\documentclass[letterpaper,twocolumn,10pt]{article}
\usepackage{usenix}

\usepackage{tikz}
\usepackage{amsmath}

\usepackage{filecontents}

\usepackage{url}
\usepackage{xspace}
\usepackage{caption}
\usepackage{booktabs}

\newcommand{\name}[0]{Computron\xspace}
\newcommand{\repo}[0]{\url{https://github.com/dlzou/computron}}

\begin{document}

\date{}

\title{\name: Serving Distributed Deep Learning Models \\ with Model Parallel Swapping}

\author{
\rm{Daniel Zou$^\text{1, *}$ \enskip
    Xinchen Jin$^\text{2}$ \enskip
    Xueyang Yu$^\text{2}$ \enskip
    Hao Zhang$^\text{3}$ \enskip
    James Demmel$^\text{1}$ \enskip}
\\
{$^{\text{1}}$UC Berkeley\enskip $^{\text{2}}$ShanghaiTech University\enskip $^{\text{3}}$UC San Diego\enskip}
} 

\maketitle

{\let\thefootnote\relax\footnote{{$^*$dlzou@berkeley.edu}}}
\begin{abstract}

Many of the most performant deep learning models today in fields like language and image understanding are fine-tuned models that contain billions of parameters. In anticipation of workloads that involve serving many of such large models to handle different tasks, we develop \name, a system that uses memory swapping to serve multiple distributed models on a shared GPU cluster. \name implements a model parallel swapping design that takes advantage of the aggregate CPU-GPU link bandwidth of a cluster to speed up model parameter transfers. This design makes swapping large models feasible and can improve resource utilization. We demonstrate that \name successfully parallelizes model swapping on multiple GPUs, and we test it on randomized workloads to show how it can tolerate real world variability factors like burstiness and skewed request rates. \name's source code is available at \repo.

\end{abstract}

\section{Introduction}
\label{sec:intro}

In recent years, researchers and practitioners have dramatically improved the performance of deep learning models, particularly large language models (LLMs), using two techniques: massive parameterization and fine-tuning. Many pre-trained models with billions of parameters have been released, and each of them is being customized for a myriad of tasks through fine-tuned variants. In a plausible scenario, organizations would host many of these large models, each similar in architecture and size but tuned to some specific task, to serve the needs of their internal personnel and external users.

The usual way to serve a model using GPUs is to keep all of its parameters in GPU memory so that inference runs directly on the accelerator device. When a model is too large to fit in a single GPU's memory, a common technique is to distribute it to multiple GPUs through model parallelism. The amount of memory onboard each GPU is limited, so an organization would need to purchase many GPUs to serve all of its models, which could be quite expensive. Worse, the costly hardware is underutilized when some models receive requests at low or irregular rates.

Among existing ML serving systems, some such as AlpaServe \cite{li2023alpaserve} and Energon-AI \cite{du2022energonai} employ model parallelism to serve large models in a distributed fashion. There are also systems like Clockwork \cite{gujarati2020serving} that use memory swapping to overcome the limitation of GPU memory and improve utilization. 

In this paper, we present \name, a prototype serving system that unifies model parallelism and swapping. \name makes it possible to serve multiple large distributed models that, in total, can exceed the memory capacity of the GPU cluster they share. In terms of usability, \name supports Colossal-AI's \cite{li2022colossalai} functionality for easy model parallelism during model development, and it integrates with asynchronous Python frameworks for service deployment. We discuss several ordering and synchronization problems that constrain the design of such a system, and we explain how our design solves these problems to achieve parallelized swapping on multiple GPUs. We evaluate \name in two ways. First, we isolate the swapping component to demonstrate that model parallel swapping does in fact reduce the time taken to load a distributed model into GPU memory. Second, we test \name under more realistic conditions on randomly generated workloads that simulate conditions where requests may be bursty and skewed to a subset of models.

\section{Background and Problem}
\label{sec:background}

Deep learning is being rapidly adopted in countless business and scientific applications. In many of these applications where some form of service deployment is involved, the serving system is a crucial component of the deep learning workflow. These systems generally operate in a request-response paradigm by listening to inference requests, running the requested model on inputs on specialized hardware such as GPUs, then responding with the output. The design of such a system revolves around the tradeoff of reducing latency so that end users experience less waiting versus increasing efficiency to save operational cost, all without compromising model accuracy.

Recent research in language models has popularized the practice of fine-tuning, where either a portion or all of a pre-trained model's parameters are trained on new data from a specific task in order to achieve higher accuracy. For example, the pre-trained BERT \cite{devlin2019bert} model can be fine-tuned to a variety of language understanding tasks---from text classification to part-of-speech tagging to natural language inference---just by retraining the last layers, and GPT-3 \cite{brown2020language} has been fine-tuned on human feedback so that the resulting InstructGPT \cite{ouyang2022training} model can better align to user intentions. Fine-tuning generally involves none or minor modifications of the model architecture. As fine-tuned models become commonplace, so will workloads that involve serving multiple models with highly similar memory footprints and access patterns.

A second trend spearheaded by the language modeling community has been to increase model size in pursuit of better accuracy and generalizability. At the extreme, Megatron-Turing NLG \cite{smith2022using} contains 530 billion parameters, and the comparatively space-efficient LLaMA \cite{touvron2023llama} still contains up to 65 billion parameters. On Chatbot Arena \cite{zheng2023judging}, a platform where humans rate the quality of language model outputs, most models with competitive performance have at least 6 billion parameters. 

Serving multiple instances of such large models would exceed the memory capacity of all consumer GPUs and many high-end ones. The standard solution is to distribute large models across multiple GPUs using model parallelism. Two forms, tensor parallelism (TP) and pipeline parallelism (PP), are well-studied and commonly used in training workloads \cite{narayanan2021efficient}, but they are just recently beginning to see use in serving systems as well. Even when all models can fit into aggregate GPU memory, the AlpaServe \cite{li2023alpaserve} team found that there is still reason to use model parallelism in a serving system because it can reduce latency in real world workloads.

Another challenge of serving deep learning models is that real world serving workloads are often unpredictable. Request arrival distributions may be bursty. Furthermore, across multiple models, request rates may be skewed---some models may receive a lot more requests than others---and the rates may shift over time. Hosting all of these models in GPU memory leads to underutilization in the face of dynamic request patterns, as resources are over-provisioned to models with low request rates. The resource allocation imbalance is worse for localized services that expect irregular traffic, and the hardware cost is higher when larger models distributed across multiple devices are involved. Both of these factors act against the interests of smaller organizations. 

We survey a number of prior works and find that while there are many designs from which to take inspiration, to the best of our knowledge, no single system addresses both the model parallelism and the resource utilization challenges that we have outlined. AlpaServe \cite{li2023alpaserve} and Energon-AI \cite{du2022energonai} are capable of serving large models using model parallelism, but they host all models on GPUs following some form of static assignment and are thus bounded by the amount of available GPU memory. Clockwork \cite{gujarati2020serving} serves many deep learning models on a limited number of GPUs by swapping models between CPU and GPU memory, which works well when the models are on the order of millions of parameters or less. However, this approach is not suited for larger models with billions of parameters that can take several seconds to transfer. ZeRO-Inference \cite{aminabadi2022deepspeed} parallelizes parameter transfers across GPUs in order to multiply CPU-GPU bandwidth, but it is meant for individual layers within a single massive model.

\section{Design}
\label{sec:design}

In \S\ref{sec:background}, we provide motivation for the problem of serving multiple distributed deep learning models, and we identify the key issue of resource underutilization when handling bursty, skewed requests. To deal with these challenges, we propose a serving system that co-locates multiple models on the same cluster of GPUs and dynamically swaps distributed model parameters between CPU and GPU memories. Active models are swapped into GPU memory so that requests can be served quickly, while unused models are swapped out to reduce the unnecessary consumption of resources.

The underlying technique of offloading unused models to CPU memory has already been applied by state-of-the-art serving systems like Clockwork \cite{gujarati2020serving} to great effect. Our particular approach is comparable to demand paging in the context of virtual memory management; when a model is requested whose parameters do not currently reside in GPU memory, a replacement policy is used to pick another model to swap out, and then the requested model is swapped in. Like other systems, we assume the existence of large CPU memory to hold unused model parameters. More sophisticated fetching algorithms can be used here instead, but are beyond the scope of this paper.

Inspired by prior works, we seek to investigate whether model parallelism can also be used to reduce the latency of model swapping. We hypothesize that on systems where GPUs have independent PCIe links to the CPU, by specifying higher degrees of TP and PP to distribute model shards across more GPUs, model parameter shards can be loaded in parallel to take advantage of greater aggregate link bandwidth between the CPU and GPUs. A similar optimization is used by ZeRO-Inference. Should this prove true in practice, it would become feasible to perform dynamic swapping while serving large models that share a group of devices, just like smaller models. However, a number of design considerations arise when we put our hypothesis to the test.

\subsection{Architecture}
\label{subsec:architecture}

\begin{figure}
    \centering
    \includegraphics[width=\columnwidth]{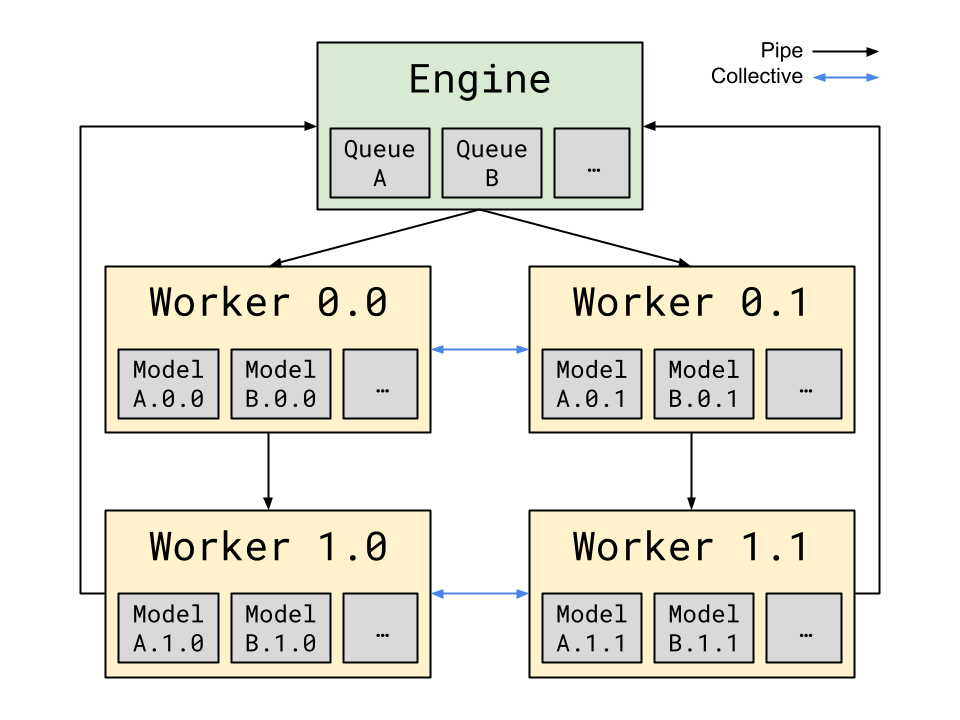}
    \caption{\name architecture for $TP = 2$, $PP = 2$. One worker is launched per GPU. Two models, A and B, are co-located in the same parallel configuration.}
    \label{fig:architecture}
\end{figure}

Our system uses an engine-worker architecture to manage multiple distributed model instances at the same time. The centralized engine receives, queues, and waits for the completion of all model requests. Workers are launched per GPU in accordance with a user-provided parallel configuration (TP and PP dimensions) to manage shards of model parameters. Because we assume all models have a similar size and architecture, we make the simplification to co-locate each distributed model instance onto the GPU cluster using the same configuration. Fig. \ref{fig:architecture} gives an example of the architecture for a $TP = 2$, $PP = 2$ configuration.

When the engine receives a request for some model, it pushes the request object along with a timestamp into a queue specifically for that model. Concurrently, the engine repeatedly picks a queue to pop oldest request objects, then packs and submits them to workers in the first pipeline stage as a single batch entry. Workers at each pipeline stage evaluate batch entries in submitted order, up to the last stage, at which point last stage workers send the batch output back to the engine. PP communication occurs through FIFO pipes, while TP communication is done through distributed collectives, as represented by the arrows in Fig. \ref{fig:architecture}.

\subsection{Model Parallel Swapping}
\label{subsec:swapping}

\begin{figure}
    \centering
    \includegraphics[width=\columnwidth]{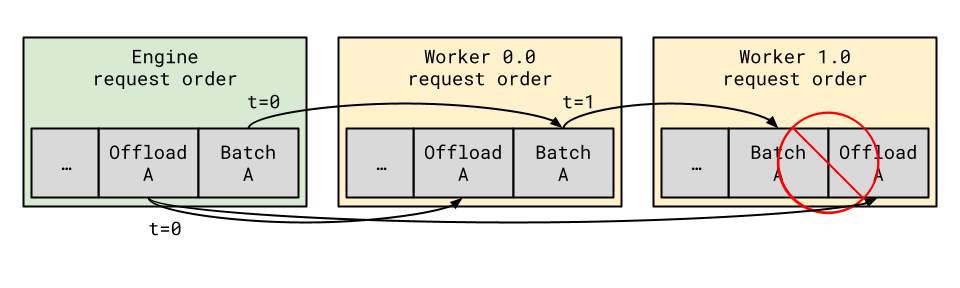}
    \caption{Broadcasted load entry violates load dependency.}
    \label{fig:challenge1}
\end{figure}

\begin{figure}
    \centering
    \includegraphics[width=\columnwidth]{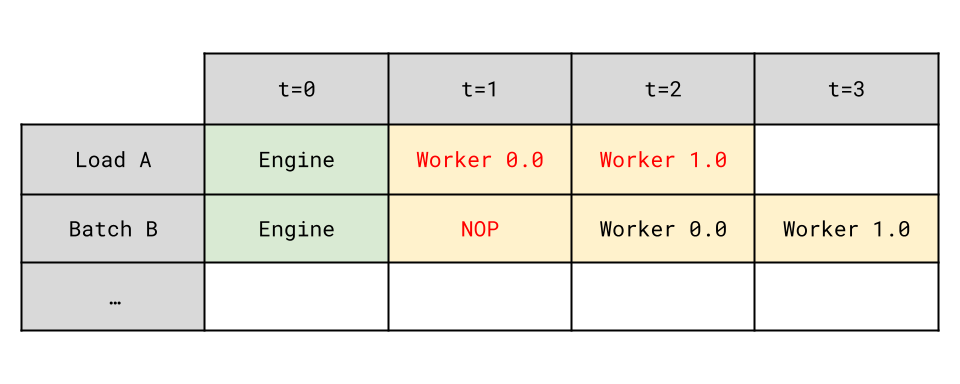}
    \caption{Synchronous load entry reduces loading parallelism and causes unnecessary blocking.}
    \label{fig:challenge2}
\end{figure}

In our design, the responsibility of making swapping decisions is delegated to the engine, so in addition to submitting batch entries, the engine can initiate another type of action through what we refer to as load entries. A load entry commands a worker to either load or offload the parameters of an instance. 

Challenges arise when designing how load entries should be submitted to and processed by distributed workers. A model can only be evaluated on a batch entry after the model's parameters are loaded into GPU memory, so as the engine schedules batch and load entries in some order it deems correct, workers must respect the load dependencies of that schedule. Furthermore, data dependencies between adjacent pipeline stage workers delay when later stage workers receive batch entries, ruling out certain designs like broadcasting the load entry, as illustrated in Fig. \ref{fig:challenge1}. These load and data dependencies are resolved if workers synchronously process load entries in pipeline order just like batch entries, but this naive solution has two issues: a batch entry to some model is unnecessarily blocked by load entries to another unrelated model, and no loading parallelism is achieved by workers of different stages in the same pipeline, as shown in Fig. \ref{fig:challenge2}.

\begin{figure}
    \centering
    \includegraphics[width=\columnwidth]{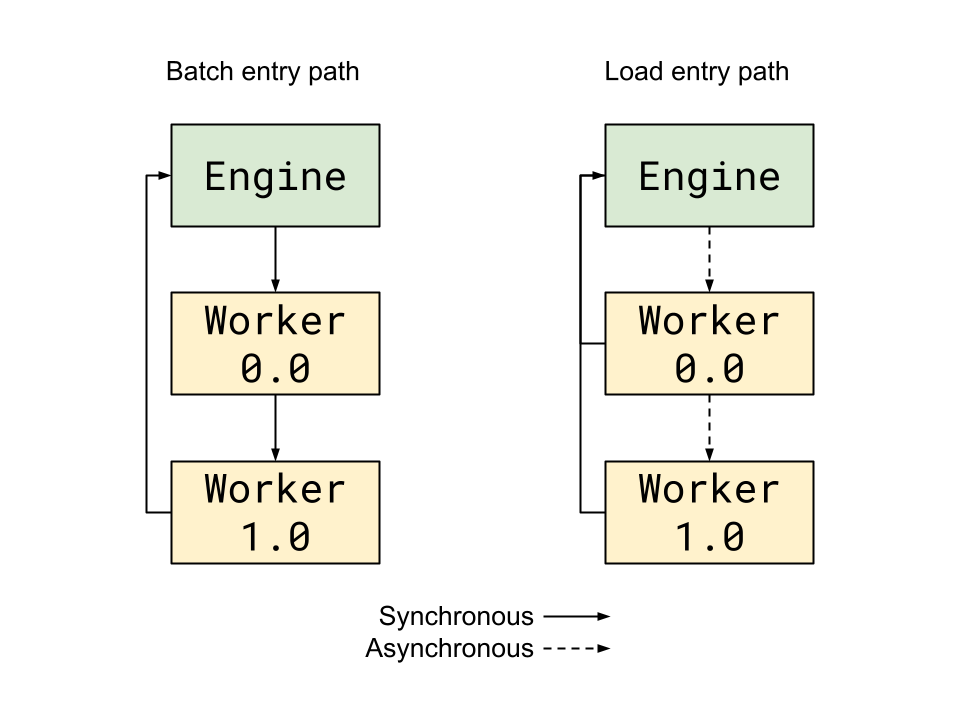}
    \caption{Comparison of how batch entries and load entries are processed in a linear worker pipeline.}
    \label{fig:entries}
\end{figure}

We propose an asynchronous mechanism for handling load entries that mitigates these issues. After being submitted by the engine, load entries are pipelined through worker stages just like batch entries, but a worker does not wait for loading to complete before passing the load entry forward to the next stage. This can be done using the stream feature of the CUDA programming model. On top of the default CUDA stream that executes kernels for model inference, each worker launches two additional streams to run loading and offloading operations concurrently. A load entry is completed when every worker finishes loading/offloading and sends a response back to the engine. The engine is responsible for avoiding load dependency violations by making sure batch entries for a model are submitted to workers only after that model has been fully loaded. This design allows a later batch entry to proceed without waiting for a previous load entry involving another model to complete, and this also enables workers of different stages to load shards of a model's parameters in parallel. The paths of batch and load entries through one branch of the system pipeline are depicted in Fig. \ref{fig:entries}.

One more detail is the use of pinned memory. CUDA requires the CPU-side data buffer to be in page-locked memory during CPU-GPU data transfers to prevent interruptions caused by paging. Data objects on CPU are stored in paged memory by default, so data transfers would incur an extra copy on the CPU side from paged memory to page-locked memory. We eliminate this extra data movement by making sure that when a model is offloaded, the parameters are kept pinned in CPU memory.

\section{Implementation}
\label{sec:implementation}

We build \name as a serving system that supports model parallel swapping based on the considerations presented in \S\ref{sec:design}. We borrow some components from Energon-AI \cite{du2022energonai}, such as the RPC-based FIFO pipe implementation used for communication between pipelined worker stages. Just like Energon-AI, \name is compatible with Colossal-AI \cite{li2022colossalai} functionality, meaning that users can easily incorporate model parallelism in their models with minimal changes to PyTorch source code.

As \name launches, Colossal-AI automatically handles setting up the context and communication groups for model parallelism, and it does so using the same configuration for each instance. The engine is implemented using Python's asyncio library, and request scheduling is done in a completely asynchronous fashion. Because of this, \name integrates with asynchronous Python web frameworks such as FastAPI. Requests are scheduled in batches based on the oldest timestamp, and model swapping uses an LRU replacement policy.

\section{Evaluation}
\label{sec:eval}

We design two separate sets of experiments in order to characterize \name's performance. In the first set of experiments, we intentionally induce the worst case for handling each request and measure how the time to swap models scales with model parallelism. In the second set, we generate simulated request workloads using a random arrival process to study how \name handles more realistic scenarios.

We conduct experiments on a single GPU node of the Perlmutter supercomputer managed by NERSC. The GPU node has one AMD EPYC 7763 CPU and four NVIDIA A100 GPUs, each connected to the CPU through a PCIe 4.0 x16 link \cite{nersc2022arch}.

\subsection{Swapping Latency}
\label{subsec:eval-swap}

In \S\ref{sec:design}, we hypothesize that model parallelism linearly decreases the time taken to load and offload model parameters between CPU and GPU memory. To check this hypothesis, we design an experiment that forces the worst case scenario where each request must perform a swap. We launch models concurrently and configure the engine to only allow one model to reside in GPU memory at any given time. We then send alternating blocking requests to the two models and measure the times taken to swap and execute a model at each request. The model size is fixed in order to test the strong scaling properties of CPU-GPU swapping. The model chosen is OPT-13B \cite{zhang2022opt}, an open-source pre-trained transformer language model released by Meta AI. Using half-precision floats, OPT-13B has a memory footprint of about 24 GB. This model is chosen because it can fit into the memory of a single A100 GPU, which serves as a baseline for swapping time.

Before running the experiment, we estimate the lower bound for how long swapping should take for comparison. Each CPU-GPU link has a bandwidth of 32 GB/s, so a single GPU is expected to load or offload an OPT-13B model instance in $24/32 = 0.75$ seconds. On our test system, aggregate CPU-GPU bandwidth increases linearly with the number of GPUs, so as the model is distributed to more GPUs using either TP or PP, the swapping time is expected to inversely decrease. Swapping time includes both the offloading of one model and the loading of another, and because our asynchronous implementation overlaps the two, we measure from when the offload entry is submitted to when both offload and load entries are completed.

\begin{figure}
    \centering
    \includegraphics[width=\columnwidth]{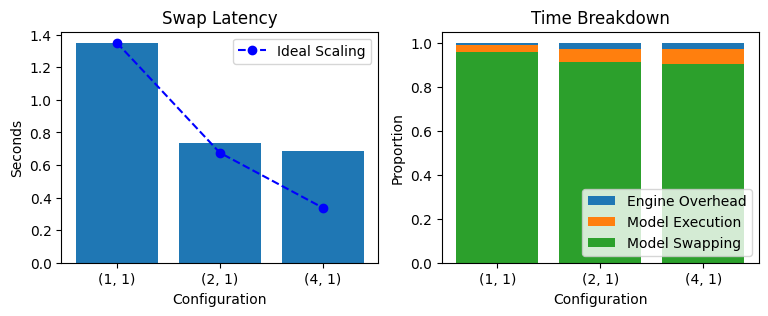}
    \caption{Swapping latency with changing TP scale.}
    \label{fig:rr-tp}
\end{figure}

With the theoretical lower bound in mind, we first run the experiment with three trials that scale the degree of model TP. We use a small input token length of 2. Fig. \ref{fig:rr-tp} visualizes the results of these trials on $TP = 1$, $TP = 2$, and $TP = 4$, all with $PP = 1$. The left plot examines how average time spent swapping scales with TP. These trials confirm that the swap latency does decrease as TP increases, as we hypothesized. However, the latency on a single GPU is noticeably higher than the lower bound, and the scaling appears to be less than linear; this difference may be explained by the alpha-beta communication model. Model parameters are transmitted not as one long stream, but as separate messages for the individual tensors. Each TP shard still contains the same number of tensors as the original model albeit smaller, so the same number of messages must be sent by each worker when loading. Taking the expression $\alpha + \beta*n$ for total latency, while message size $n$ is reduced per worker, the per message latency term $\alpha$ remains the same, leading to sublinear scaling. 

The right plot shows swapping and execution times in proportion to the end-to-end latency. From the plot, it is clear that swapping latency remains the bottleneck in all cases, but as the number of GPUs increases, the proportion of overall time spent swapping decreases; this highlights how model parallelism benefits swapping even more than execution. 

\begin{figure}
    \centering
    \includegraphics[width=\columnwidth]{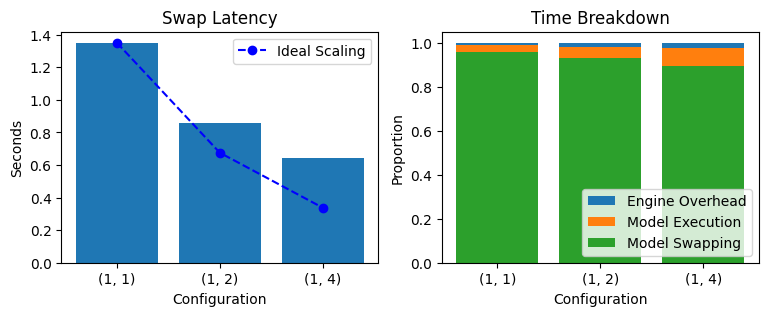}
    \caption{Swapping latency with changing PP scale.}
    \label{fig:rr-pp}
\end{figure}

Next, we run an experiment that varies PP degree between 1, 2, and 4 worker stages. Similar to the TP experiments, Fig. \ref{fig:rr-pp} shows that increasing PP also decreases the swapping latency. We postulate that in this case, sublinear scaling stems from delays as a load entry is pipelined through workers. Since workers process batch entries synchronously, load entries, despite being asynchronous, must still wait for their turn.

\begin{figure}
    \centering
    \includegraphics[width=\columnwidth]{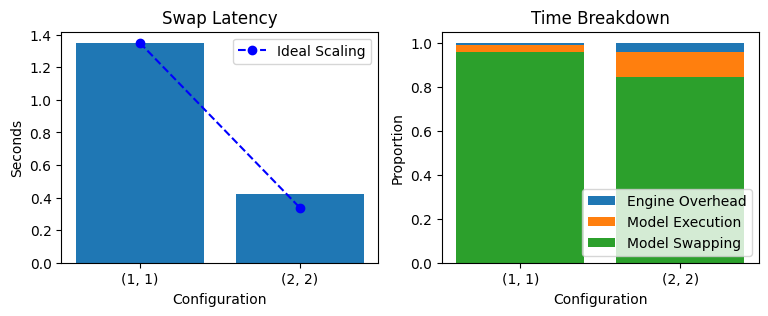}
    \caption{Swapping latency for $TP = 2$, $PP = 2$.}
    \label{fig:rr-tp-pp}
\end{figure}

TP and PP are often used together in practice, so we ran an additional trial with $TP = 2$, $PP = 2$. From Fig. \ref{fig:rr-tp-pp}, we see that the mixed parallelism configuration has lower latency than both pure TP and pure PP for the same number of workers, and it in fact approaches the ideal scaling target. The positive effect of mixing parallelism may be because the previously described overheads in the TP case and in the PP case are lessened at smaller degrees.

\subsection{Simulated Workloads}
\label{subsec:eval-sim}

We characterize the practical performance of our model parallel swapping design with simulated workloads of serving multiple OPT-13B models on the same cluster of four A100 GPUs. For all experiments conducted, we follow the configuration of $TP = 2$, $PP = 2$.

Each simulation trial begins with several warm up requests that are not recorded. Then, requests are sent to all models over a 30 second period, with the distribution of requests to each model following a random independent Gamma arrival process. Each request has an input token length of 8. Across simulations, we vary two parameters: the assignment of mean arrival rates to each model, and the coefficient of variation (CV) that is shared by all models. For our purposes, assigning different mean arrival rates simulates how request rates skew toward a subset of models, while CV adjusts the burstiness of requests. For instance, $CV = 4$ is a high degree of burstiness, and $Rates = (10, 1, 1)$ represents a skew toward the first model relative to the other two.

\begin{figure}
    \centering
    \includegraphics[width=\columnwidth]{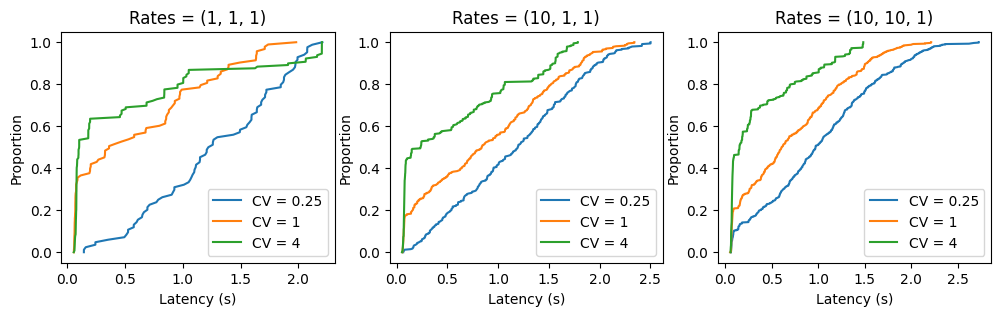}
    \caption{Latency CDFs for combinations of (Rates, CV) when serving 3 models with only 2 models in GPU memory.}
    \label{fig:sim-3-2}
\end{figure}

Our first set of simulations serves three models at once, limiting to at most two models in GPU memory at all times, and we check that GPU memory usage approximately matches the footprint of two OPT-13B models. The maximum batch size is 8. Tab. \ref{table:sim-3-2} summarizes the average end-to-end latencies for the grid of parameters we measured, with three variations in skew and three variations in CV. Fig. \ref{fig:sim-3-2} visualizes request latency CDFs of all models combined for each pair of (Rates, CV).

We observe a common pattern that as CV increases from 0.25 to 4, the latency tends to decrease, which can be seen both in the table of average latencies and in the CDF curves in each plot shifting toward the top left corner. This suggests \name performs better when request patterns are bursty. Intuitively, bursty request distributions mean higher likelihood of consecutive requests to the same model, so less model swaps occur because the engine schedules the oldest request first with LRU replacement. 

In the three-model simulation, changing the skew of request rates only marginally increases the maximum latency, and in general has little impact on the overall latency distribution. This provides evidence that \name can tolerate workloads with imbalanced requests rates. Though our simulations only test static request rate assignments, we expect that this tolerance can also extend to dynamic scenarios where the skewness of request rates changes over time. 

\begin{table}
    \centering
    \begin{tabular}{llll}
        \toprule
        Skew            & $CV = 0.25$ & $CV = 1$ & $CV = 4$ \\ 
        \midrule
        $(1, 1, 1)$     & 1.262 & 0.606 & 0.518 \\ 
        $(10, 1, 1)$    & 1.172 & 0.886 & 0.550 \\ 
        $(10, 10, 1)$   & 1.014 & 0.716 & 0.374 \\ 
        \bottomrule
    \end{tabular}
    \caption{Average latencies for combinations of (Rates, CV) when serving 3 models with only 2 models in GPU memory.}
    \label{table:sim-3-2}
\end{table}

\begin{figure}
    \centering
    \includegraphics[width=\columnwidth]{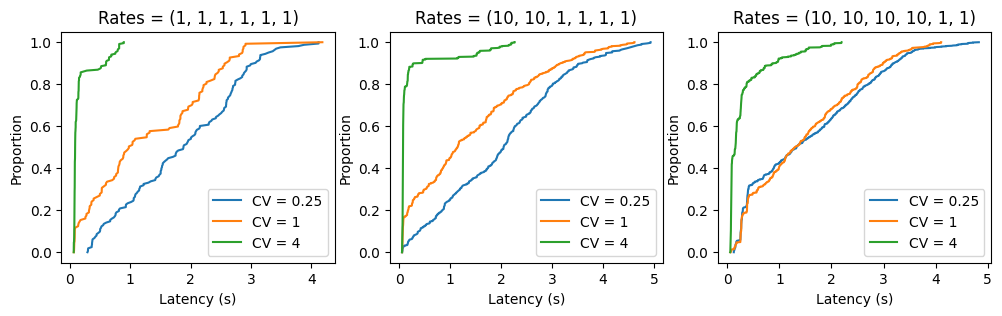}
    \caption{Latency CDFs for combinations of (Rates, CV) when serving 6 models with only 4 models in GPU memory.}
    \label{fig:sim-6-4}
\end{figure}

The second set of simulations serves six models at once, limiting to at most four models in GPU memory at all times and with the maximum batch size set to 32. The results in Fig. \ref{fig:sim-6-4} show similar patterns as the previous three-model simulations. When $CV = 4$, the latency distribution of serving six models is actually lower than serving three models on average based on Tab. \ref{table:sim-6-4}, which indicates that good resource utilization can be achieved when requests are bursty. On the other hand, latencies of lower CV trials are scaled by approximately a factor of two. A possible explanation for this is that in lower CV trials, GPUs have already been maximally utilized conditioned on the request distribution and scheduling order, so doubling the workload leads to doubled latency. 

Looking at the bigger picture, both the three-model and the six-model simulations reveal that many requests take longer than the isolated latency measurements from \S\ref{subsec:eval-swap} would suggest. Two possible causes for this are the scheduling algorithm and the choice of maximum batch size. Our simple oldest-request-first scheduling algorithm overlooks global information that may be used to reduce average latency. The maximum batch size trades off between the rate at which a model's request queue is drained and the compute time of that batch, and it may have some interactions with the request arrival distributions. We defer more thorough investigation of these effects to future work.

\begin{table}
    \centering
    \begin{tabular}{llll}
        \toprule
        Skew            & $CV = 0.25$ & $CV = 1$ & $CV = 4$ \\ 
        \midrule
        $(1, 1, 1, 1, 1, 1)$    & 1.847 & 1.282 & 0.174 \\ 
        $(10, 10, 1, 1, 1, 1)$  & 2.017 & 1.413 & 0.229 \\ 
        $(10, 10, 10, 10, 1, 1)$& 1.535 & 1.470 & 0.312 \\ 
        \bottomrule
    \end{tabular}
    \caption{Average latencies for combinations of (Rates, CV) when serving 6 models with only 4 models in GPU memory.}
    \label{table:sim-6-4}
\end{table}

\section{Conclusion}
\label{sec:conclusion}

We design and implement \name, a system that is capable of serving multiple deep learning models with billions of parameters on the same cluster of GPUs, and it can exceed aggregate GPU memory capacity through model parallel swapping. In isolated tests, we demonstrate that our design takes advantage of both TP and PP to speed up the swapping of distributed models. Simulated random workloads show that our system can tolerate bursty and skewed request patterns. These features enable organizations to efficiently serve many cutting-edge large models for different tasks when compute resources are limited.

An optimization that may significantly improve serving performance is to speculatively load or offload models. In real world scenarios, requests to different models are often not independent processes, but instead have predictable patterns, such as the same model being requested many times consecutively to generate a sequence, a subset of models often being requested in some fixed order, or a model being more frequently requested at a certain time of day. More sophisticated load scheduling algorithms with predictive capabilities can drastically reduce the number of on-demand swaps, and by extension, serving latency.

A problem that has not been resolved in this work is handling models with different sizes, and even different model parallelism configurations. Our system currently assumes that every model instance is evenly distributed across the cluster in the same way and with the same memory footprint. Removing that assumption brings many complexities such as the decision problem of what to load/offload when swapping and whether workers should handle each model differently.

\bibliographystyle{plain}
\bibliography{computron}

\end{document}